\documentclass[aps,prl]{revtex4-1}

\begin{document}
\author{Wei-Min Sun$^{1,4}$, Xiang-Song Chen$^{2,1,4}$, Xiao-Fu L$\ddot{\mathrm{u}}$$^3$, Fan Wang$^{1,4}$}
\address{$^1$Department of Physics, Nanjing University, CPNPC, Nanjing
210093, China}
\address{$^2$Department of Physics, Huazhong University
of Science and Technology, Wuhan 430074, China}
\address{$^3$Department of Physics, Sichuan University, Chengdu 610064, China}
\address{$^4$Kavli Institute for Theoretical Physics China, CAS, Beijing
100190, China}
\title{Gauge invariant hydrogen atom Hamiltonian}
\begin{abstract}
For quantum mechanics of a charged particle in a classical external
electromagnetic field, there is an apparent puzzle that the matrix
element of the canonical momentum and Hamiltonian operators is gauge
dependent. A resolution to this puzzle is recently provided by us in
\cite{ChenWang}. Based on the separation of the electromagnetic
potential into pure gauge and gauge invariant parts, we have
proposed a new set of momentum and Hamiltonian operators which
satisfy both the requirement of gauge invariance and the relevant
commutation relations. In this paper we report a check for the case
of the hydrogen atom problem: Starting from the Hamiltonian of the
coupled electron, proton and electromagnetic field, under the
infinite proton mass approximation, we derive the gauge invariant
hydrogen atom Hamiltonian and verify explicitly that this
Hamiltonian is different from the Dirac Hamiltonian, which is the
time translation generator of the system. The gauge invariant
Hamiltonian is the energy operator, whose eigenvalue is the energy
of the hydrogen atom. It is generally time-dependent. In this case,
one can solve the energy eigenvalue equation at any specific instant
of time. It is shown that the energy eigenvalues are gauge
independent, and by suitably choosing the phase factor of the
time-dependent eigenfunction, one can ensure that the time-dependent
eigenfunction satisfies the Dirac equation.


\end{abstract}

\maketitle
 \section{I.~INTRODUCTION}
In quantum mechanics, the momentum and Hamiltonian are the
fundamental physical quantities of a system. The
momentum/Hamiltonian operators are the space/time translation
generators of the system. The momentum operators satisfy the
canonical momentum commutation relations.

Gauge invariance has been recognized as a first principle through
the development of the standard model. For a charged particle in a
classical external electromagnetic field, the gauge invariance
principle requires that the matrix element of any physical
observable of the system should be gauge invariant. However, there
are apparent puzzles concerning the momentum and Hamiltonian
operators of the charged particle. For example, the expectation
value of the Hamiltonian of the hydrogen atom is gauge dependent
under a time dependent gauge transformation \cite{Goldman}. The
matrix element of the canonical momentum operator is also gauge
dependent.

A resolution to this puzzle has been recently given by us in
\cite{ChenWang}. The key idea of our resolution is to separate the
electromagnetic potential into pure gauge and gauge invariant parts.
Based on this separation, we have proposed a new set of momentum and
Hamiltonian operators which satisfy both the requirement of gauge
invariance and the relevant commutation relations.

In the present paper, following our previous work, we will show an
explicit check for the case of hydrogen atom problem: Starting from
the total Hamiltonian of the coupled electron, proton and
electromagnetic field, under the infinite proton mass approximation,
we derive the gauge invariant Hamiltonian of the hydrogen atom and
verify the difference between this Hamiltonian and the time
translation generator, the Dirac Hamiltonian.

In section II, we describe the conflict between gauge invariance and
canonical quantization of the momentum and Hamiltonian operators for
a charged particle in an external electromagnetic field and our
resolution to this problem. In section III, we give the explicit
check for the case of hydrogen atom problem. The last section
provides a summary. \vskip 0.3cm

\section{II.~GAUGE INVARIANCE AND CANONICAL QUANTIZATION OF THE
MOMENTUM AND HAMILTONIAN OPERATORS}

In classical mechanics, the canonical momentum and Hamiltonian for a
nonrelativistic particle in an external electromagnetic field
$A^\mu$ are
\begin{equation}
{\vec p}=m{\vec v}-e{\vec A}£¬~~~~~~~H=\frac{1}{2m}({\vec p}+e{\vec
A})^2-eA^0£¬
\end{equation}
where the charge of the particle is $-e$. These two dynamical
variables are gauge dependent and so are not observables in
classical gauge theory. After quantization, the momentum ${\vec p}$
is quantized as ${\vec p}=-i{\vec \nabla}$ (in coordinate
representation), irrespective of which gauge is chosen, even though
the classical canonical momentum is gauge dependent. The Hamiltonian
is quantized by replacing ${\vec p}$ with $-i{\vec \nabla}$. The
quantized momentum operators satisfy the canonical momentum
commutation relations $[p^i,p^j]=0$.

After a gauge transformation£º
\begin{equation}
\psi'=e^{-ief(x)}\psi,~~~~~~A'^\mu=A^\mu-\partial^\mu f(x),
\end{equation}
the expectation value of the above operators transform as follows
\begin{equation}
\langle \psi'|{\vec p}|\psi'\rangle=\langle \psi|{\vec p}|\psi
\rangle-e\langle \psi|{\vec \nabla}f|\psi\rangle,~~~~~~ \langle
\psi'|H'|\psi'\rangle=\langle \psi|H|\psi\rangle+e\langle
\psi|\frac{\partial f}{\partial t}|\psi \rangle.
\end{equation}
The expectation values of these two operators are gauge dependent.
Therefore these expectation values are not measurable and hence
these operators are not observables.

The same problem also exists in relativistic quantum mechanics. The
gauge dependence of the expectation value of the Hamiltonian of the
electron in an external electromagnetic field under a time dependent
gauge transformation was discussed in \cite{Goldman}.

In order to resolve the puzzle of gauge invariance of the
expectation value of canonical momentum, one introduces the gauge
invariant operator
\begin{equation}
\vec{\cal P}={\vec p}+e{\vec A}.
\end{equation}
It is easy to check that the expectation value of this operator is
gauge invariant. However, the commutators between the components of
$\vec{\cal P}$ are
\begin{equation}
[{\cal P}^i,{\cal P}^j]=ie(\partial^i A^j-\partial^j A^i)=ieF^{ij},
\end{equation}
therefore $\vec{\cal P}$ does not satisfy the Lie algebra of
canonical momentum, so it cannot be the proper momentum opearator.

A resolution to this problem is given by us in \cite{ChenWang}. Our
idea is to seek a unique separation
$A^\mu=A^\mu_{pure}+A^\mu_{phys}$, with $A^\mu_{pure}$ a pure gauge
term having the same transformation property as the full $A^\mu$ and
giving null field strength, and $A^\mu_{phys}$ a physical term which
is gauge invariant. The condition that $A_{pure}$ gives null field
strength reads
\begin{equation} \label{nullfieldstrength}
 \partial^\mu A^\nu_ {pure}-\partial^\nu A^\mu_{pure} =0.
\end{equation}
 This equation cannot fix $A_{pure}$ uniquely. One needs to find an
 additional condition to fix it.
 The spatial part of  Eq. (\ref{nullfieldstrength}) is
 \begin{equation}
 {\vec \nabla} \times {\vec A}_{pure}=0.
 \end{equation}
 %This equation can be expressed equivalently in the form
 %\begin{equation} \label{curl}
 %{\vec \nabla} \times {\vec A}_{phys} =  {\vec \nabla} \times {\vec
 %A}.
 %\end{equation}
A natural choice of the additional condition is
\begin{equation} \label{div}
 {\vec \nabla} \cdot {\vec A}_{phys}=0.
\end{equation}
That is, ${\vec A}_{phys}$ and ${\vec A}_{pure}$ are the transverse
component ${\vec A}_\bot$ and longitudinal component ${\vec
A}_{//}$, respectively. The time component $A^0$ can be decomposed
in the same manner. From the condition $F^{i0}_{pure}=0$, one
obtains
\begin{equation}
\partial_i A^0_{phys}=\partial_i A^0 +\partial_t(A^i-A^i_{phys}).
\end{equation}
From the above equation one can derive
\begin{equation}
A^0_{phys}=\int_{-\infty}^{x}d x^i (\partial_i A^0+\partial_t A^i-\partial_t A^i_{phys})
\end{equation}
(Here no summation over the index $i$ is implied).

%one can write
%\begin{equation}
%{\vec A}_{phys}=-{\vec \nabla}\times \frac{{\vec \nabla} \times {\vec A}}{\nabla^2}
%={\vec A}-{\vec \nabla}\frac{{\vec \nabla}\cdot {\vec A}}{\nabla^2}={\vec A}_{\bot},~~~~
%{\vec A}_{pure}= {\vec \nabla}\frac{{\vec \nabla}\cdot {\vec A}}{\nabla^2}={\vec A}_{//},
%\end{equation}
%i.e., ${\vec A}_{phys}$ and ${\vec A}_{pure}$ are the transverse
%and longitudinal component of ${\vec A}$, respectively. Since $A^\mu_{pure}$ should be a pure gauge term,
%one has
%\begin{eqnarray}
%A^0_{pure}&=&-\frac{\partial}{\partial t}\frac{{\vec \nabla}\cdot
%{\vec A}}{\nabla^2}, ~~A^0_{phys}=A^0-A^0_{pure}.
%\end{eqnarray}

Based on the gauge field decomposition above, we introduce another
momentum operator
\begin{equation}\label{momentum}
{\vec p}_{pure}={\vec p}+e{\vec A}_{pure}.
\end{equation}
This operator satifies both the requirement of gauge invariance
(because ${\vec A}_{pure}$ has the same gauge transformation
property as the full ${\vec A}$) and the the Lie algebra for the
canonical momentum (because $A^\mu_{pure}$ gives null field
strength).

The long-standing puzzle of the gauge non-invariance of the
expectation value of the Hamiltonian \cite{Goldman} can be solved in
the same manner. For nonrelativistic quantum mechanics, we define a
new Hamiltonian
\begin{eqnarray}
H &=& \frac{({\vec p}+e{\vec A})^2}{2m}-eA^0+eA^0_{pure} \nonumber \\
&=& \frac{({\vec p}+e{\vec A})^2}{2m} -eA^0_{phys}.
\end{eqnarray}
The term $eA^0_{pure}$ is a pure gauge term which cancels the
unphysical energy appearing in $-eA^0$ and then guarantees that the
expectation value of this Hamiltonian is gauge invariant. It is a
direct extension of Eq. (\ref{momentum}) to the zeroth momentum
component.

Therefore, for a charged particle in a classical external
electromagnetic field, the gauge invariant momentum/Hamiltonian
operator is not the space/time translation generator of the system.
The gauge invariant momentum and Hamiltonian are observables,
whereas the space and time translation generators are not.

The Dirac Hamiltonian has the same unphysical energy part which must
be canceled in the same manner as for the Schroedinger Hamiltonian
(The distinction between the gauge invariant Hamiltonian and the
time translation generator in this case was also pointed out by Kobe
and Yang in \cite{Kobe}). In the rest part of this paper, we shall
do a check for the case of the hydrogen atom problem: Starting from
the Hamiltonian of the coupled electron, proton and electromagnetic
field, under the infinite proton mass approximation, we derive the
gauge invariant hydrogen atom Hamiltonian and verify the difference
between this Hamiltonian and the time translation generator, the
Dirac Hamiltonian.

\section{III.~DERIVATION OF GAUGE INVARIANT HYDROGEN ATOM
HAMILTONIAN}

Let us start from the coupled-field Lagrangian for electron,~proton
and electromagnetic field
\begin{equation}\label{Lagrangian}
{\cal L}=-\frac{1}{4}F_{\mu\nu} F^{\mu\nu}+{\bar \psi}_e(i\gamma^\mu
D_\mu^{(e)}-m)\psi_e+{\bar \psi}_p(i\gamma^\mu
D_\mu^{(p)}-M)\psi_p~~~~~~~~~~~~~~
\end{equation}
with $D_\mu^{(e)}=\partial_\mu-ie A_\mu$ and
$D_\mu^{(p)}=\partial_\mu+ie A_\mu$ (here $e$ is the charge of the
proton). From this Lagrangian one can derive the total energy of the
system
\begin{equation}\label{totalenergy}
H=\int d^3 x \Big[ \psi^\dagger_e(i {\vec \alpha}\cdot {\vec
D}^{(e)}+\beta m)\psi_e+\psi^\dagger_p(i {\vec \alpha}\cdot {\vec
D}^{(p)}+\beta M)\psi_p+\frac{1}{2}(E^2+B^2) \Big].
\end{equation}

To proceed, we decompose the gauge potential ${\vec A}$ in terms of
its transverse and longitudinal parts:
\[ {\vec A}={\vec A}_{phys} +{\vec A}_{pure}={\vec A}_\bot+{\vec A}_{//}.
\]
Then one has
\begin{eqnarray}\label{decompE}
{\vec E}&=&-{\vec \nabla}A^0-\frac{\partial}{\partial t}{\vec
A}_{pure}-\frac{\partial}{\partial t}{\vec A}_{phys} \nonumber \\
&=& -{\vec \nabla} A^0_{phys}-{\vec \nabla}A^0_{pure}-\frac{\partial}{\partial t} {\vec A}_{pure}-\frac{\partial}{\partial t}{\vec A}_{phys}
\nonumber \\
&=&-{\vec \nabla}A^0_{phys}-\frac{\partial}{\partial
t}{\vec A}_{phys} \nonumber \\
&=& -{\vec \nabla}A^0_{phys} -\frac{\partial}{\partial t}{\vec A}_\bot \nonumber
\\
&\equiv& {\vec E}_{//}+{\vec E}_\bot,
\end{eqnarray}
where we have used the condition that $A^\mu_{pure}$ gives null field strength.
The total electromagnetic field energy then separates into two terms
\begin{equation}\label{totalEMenergy}
\frac{1}{2}\int d^3 x (E^2+B^2)=\frac{1}{2}\int d^3 x
E_{//}^2+\frac{1}{2}\int d^3 x (E_\bot^2+B^2),
\end{equation}
where the cross term ${\vec E}_{//}\cdot {\vec E}_\bot$ vanishes by
an integration by parts. The first term on the right of
(\ref{totalEMenergy}) is the total energy associated with the
Coulomb field. In fact, from Gauss law ${\vec \nabla}\cdot {\vec
E}=e\psi_p^\dagger \psi_p -e \psi_e^\dagger \psi_e=\rho_p+\rho_e$
one has
\begin{equation}
{\vec \nabla}\cdot {\vec E}_{//}=-\nabla^2 A^0_{phys}=\rho_p+\rho_e.
\end{equation}
From the above equation one can obtain $A^0_{phys}$:
\begin{equation}
A^0_{phys}=-\frac{1}{\nabla^2}(\rho_p+\rho_e).
\end{equation}
One then has
\begin{eqnarray}
\frac{1}{2}\int d^3 x  E_{//}^2 &=& -\frac{1}{2}\int d^3x {\vec
E}_{//}\cdot {\vec \nabla}A^0_{phys} \nonumber \\
&=& \frac{1}{2}\int d^3x {\vec \nabla} \cdot {\vec E}_{//}A^0_{phys}
\nonumber \\
&=& -\frac{1}{2}\int d^3 x
(\rho_p+\rho_e)\frac{1}{\nabla^2}(\rho_p+\rho_e) \nonumber \\
&=& \frac{1}{4\pi}\int d^3 x d^3 y \rho_e({\vec x},t)\frac{1}{|{\vec
x}-{\vec y}|}\rho_p({\vec y},t) +\frac{1}{8\pi}\int d^3 x d^3 y
\rho_e({\vec x},t)\frac{1}{|{\vec x}-{\vec y}|}\rho_e({\vec y},t)
\nonumber \\
 &&+
\frac{1}{8\pi}\int d^3 x d^3 y \rho_p({\vec x},t)\frac{1}{|{\vec
x}-{\vec y}|}\rho_p({\vec y},t).
\end{eqnarray}
The second term on the right of (\ref{totalEMenergy}) is the energy
of the transverse electromagnetic field. The total energy of the
system then separates into the following terms
\begin{eqnarray}\label{totalenergysep}
H&=&\int d^3 x \psi^\dagger_e(i {\vec \alpha}\cdot {\vec
D}^{(e)}+\beta m)\psi_e+\int d^3 x \psi^\dagger_p(i {\vec
\alpha}\cdot {\vec D}^{(p)}+\beta M)\psi_p \nonumber \\
&&+\frac{1}{4\pi}\int d^3 x d^3 y \rho_e({\vec x},t)\frac{1}{|{\vec
x}-{\vec y}|}\rho_p({\vec y},t) +\frac{1}{8\pi}\int d^3 x d^3 y
\rho_e({\vec x},t)\frac{1}{|{\vec x}-{\vec y}|}\rho_e({\vec y},t)
\nonumber \\
&&+\frac{1}{8\pi}\int d^3 x d^3 y \rho_p({\vec x},t)\frac{1}{|{\vec
x}-{\vec y}|}\rho_p({\vec y},t)+\frac{1}{2}\int d^3 x ({\vec
E}_\bot^2+{\vec B}^2).
\end{eqnarray}

Now let us turn to the hydrogen atom problem. Here one assumes that
the proton is infinitely massive, i.e., $M \rightarrow \infty$. In
this limit the proton plays the role of a static source situated at
a fixed point in space which one for convenience takes to be the
origin. The electromagnetic current of the proton is then
\begin{equation}\label{protoncurrent}
j_p^\mu({\vec x},t)=(\rho_p({\vec x},t),{\vec j}_p({\vec
x},t))=(e\delta^3 ({\vec x}),{\vec 0}).
\end{equation}

In order to find out the energy of the electron in the
electromagnetic field of the proton, one needs to derive the
electromagnetic potential $A^\mu$. The equation of motion for
$A^{\mu}$ is
\begin{equation}\label{Maxwell}
 \partial^2 A^\mu-\partial^\mu(\partial \cdot
A)=j_p^\mu+j_e^\mu.
\end{equation}
Using Eq.(\ref{protoncurrent}), Eq.(\ref{Maxwell}) can be written as
\begin{eqnarray}\label{Maxwelltime}
\partial^2 A^0 -\frac{\partial}{\partial t}(\frac{\partial A^0}{\partial
t}+{\vec \nabla} \cdot {\vec A})&=& -\nabla^2 A^0
-\frac{\partial}{\partial t}({\vec \nabla} \cdot {\vec A})=e\delta^3
({\vec x})+\rho_e \\
\label{Maxwellspatial1} \partial^2 {\vec A}+{\vec
\nabla}(\frac{\partial A^0}{\partial t}+{\vec \nabla} \cdot {\vec
A})&=& {\vec j}_e.
\end{eqnarray}
When solving Eqs. (\ref{Maxwelltime}) and (\ref{Maxwellspatial1}),
one needs to choose a gauge. In the following we choose the gauge
\begin{equation}\label{gauge}
{\vec \nabla}\cdot {\vec A}({\vec x},t)=\chi({\vec x},t)=\nabla^2
f({\vec x},t),
\end{equation}
where $f({\vec x},t)$ is an arbitrary function. From Eq.
(\ref{Maxwelltime}) one can derive $A^0$
\begin{equation}\label{A0solution}
A^0({\vec x},t)=\frac{e}{4\pi r}-\frac{\partial}{\partial t}f({\vec
x},t)+\frac{1}{4\pi}\int d^3 y \frac{\rho_e({\vec y},t)}{|{\vec
x}-{\vec y}|}.
\end{equation}
Substituting (\ref{A0solution}) into Eq. (\ref{Maxwellspatial1})
gives
\begin{equation}\label{Maxwellspatial2}
\partial^2 {\vec A}({\vec x},t)+{\vec \nabla}\bigg( -\partial^2 f({\vec
x},t)+\frac{1}{4\pi}\int d^3 y \frac{1}{|{\vec x}-{\vec
y}|}\frac{\partial}{\partial t}\rho_e({\vec y},t)\bigg)={\vec
j}_e({\vec x},t),
\end{equation}
whereby one can derive ${\vec A}$
\begin{equation}\label{Aisolution}
{\vec A}({\vec x},t)={\vec \nabla}f({\vec x},t)+{(\partial^2)}^{-1}
{\vec j}_e({\vec x},t)+\frac{1}{4\pi}\int d^3 y
{(\partial^2)}^{-1}\bigg( \frac{{\vec x}-{\vec y}}{|{\vec x}-{\vec
y}|^3}\frac{\partial}{\partial t}\rho_e({\vec y},t) \bigg).
\end{equation}
From Eqs.(\ref{A0solution}) and (\ref{Aisolution}) we can attribute
$\big(\frac{e}{4\pi r}-\frac{\partial f}{\partial t},{\vec
\nabla}f\big)$ to be the electromagnetic potential produced by the
static proton source in the gauge (\ref{gauge}).

Now we can isolate the energy of the electron in the electromagnetic
field of the proton from the total energy of the system Eq.
(\ref{totalenergysep}). We only need to consider terms involving the
electron field. These are $\int d^3 x \psi^\dagger_e(i {\vec
\alpha}\cdot {\vec D}^{(e)}+\beta m)\psi_e$ and $\frac{1}{4\pi}\int
d^3 x d^3 y \rho_e({\vec x},t)\frac{1}{|{\vec x}-{\vec
y}|}\rho_p({\vec y},t)$. The term $\frac{1}{8\pi}\int d^3 x d^3 y
\rho_e({\vec x},t)\frac{1}{|{\vec x}-{\vec y}|}\rho_e({\vec y},t)$
represents the electron self energy
%the Coulomb energy between the electron field and itself
and one does not need to take it into account.

The term $\int d^3 x \psi^\dagger_e(i {\vec \alpha}\cdot {\vec
D}^{(e)}+\beta m)\psi_e$ can be written as
\begin{eqnarray}\label{Diracelectron}
&&\int d^3 x \psi^\dagger_e(i {\vec \alpha}\cdot {\vec
D}^{(e)}+\beta
m)\psi_e \nonumber \\
&=& \int d^3 x \psi^\dagger_e(-i {\vec \alpha}\cdot {\vec \nabla}+e
{\vec \alpha} \cdot {\vec A} +\beta m)\psi_e \nonumber \\
&=& \int d^3 x \psi^\dagger_e(-i {\vec \alpha}\cdot {\vec \nabla}+e
{\vec \alpha} \cdot {\vec \nabla} f +\beta m)\psi_e-\int d^3 x {\vec
j}_e({\vec
x},t) \cdot {(\partial_x^2)}^{-1}{\vec j}_e({\vec x},t) \nonumber \\
&&-\frac{1}{4\pi} \int d^3 x d^3 y ~{\vec j}_e({\vec x},t) \cdot
{(\partial_x^2)}^{-1} \bigg(\frac{{\vec x}-{\vec y}}{|{\vec x}-{\vec
y}|^3} \frac{\partial}{\partial t}\rho_e({\vec y},t)\bigg).
\end{eqnarray}
Here the first term on the right of Eq. (\ref{Diracelectron}) is the
kinetic energy of the electron and the ${\vec j}\cdot {\vec A}$ part
of the interaction energy of the electron with the electromagnetic
field of the proton. The second and third term on the right of Eq.
(\ref{Diracelectron}) represents self-interaction energy of the
electron field, which we do not need to consider.

The electron-proton Coulomb energy term is
\begin{eqnarray}\label{epCoulomb}
\frac{1}{4\pi}\int d^3 x d^3 y \rho_e({\vec x},t)\frac{1}{|{\vec
x}-{\vec y}|}\rho_p({\vec y},t)&=& \frac{e}{4\pi}\int d^3 x
\rho_e({\vec x},t)\frac{1}{|{\vec x}|} \nonumber \\
&=& \int d^3 x \psi_e^\dagger \big(-\frac{e^2}{4\pi r}\big)\psi_e.
\end{eqnarray}

So the energy of the electron in the electromagnetic field of the
proton is
\begin{equation}\label{electronenergy}
\int d^3 x \psi_e^\dagger \big( -i {\vec \alpha} \cdot {\vec
\nabla}+e {\vec \alpha} \cdot {\vec \nabla} f+\beta m
-\frac{e^2}{4\pi r}\big) \psi_e.
\end{equation}
From this expression one reads out the hydrogen atom Hamiltonian
\begin{eqnarray}\label{energyoperatorgeneral}
H &=&{\vec \alpha} \cdot ({\vec p}+e {\vec \nabla} f) +\beta m
-\frac{e^2}{4\pi r}.
\end{eqnarray}

On the other hand, from the Lagrangian (\ref{Lagrangian}) one can derive the equation of motion of
the electron field
\begin{equation}
(i \gamma^\mu \partial_\mu+e\gamma^\mu A_\mu-m)\psi_e =0.
\end{equation}
Substituting Eqs. (\ref{A0solution}) and (\ref{Aisolution}) into the above equation and dropping all
terms nonlinear in $\psi_e$, one obtains the linearized equation of motion of the electron field
\begin{equation}\label{linearizedEOM}
\Big(i \gamma^\mu \partial_\mu+e \gamma^0 (\frac{e}{4\pi r}-\frac{\partial f}{\partial t})
-e {\vec \gamma}\cdot {\vec \nabla} f-m \Big) \psi_e=0.
\end{equation}
Eq. (\ref{linearizedEOM}) is just the Dirac equation of an electron
in the external electromagnetic field of the proton source:
\begin{eqnarray}
i \frac{\partial}{\partial t} \psi_e &=& \Big(-i {\vec \alpha}\cdot {\vec \nabla}+e {\vec \alpha}\cdot
{\vec \nabla}f+\beta m -\frac{e^2}{4\pi r}+e \frac{\partial f}{\partial t}\Big) \psi_e \nonumber
\\
&=& H_D \psi_e,
\end{eqnarray}
where $H_D$ is the Dirac Hamiltonian which is the time translation generator. Here one notes that
the presence of the term $e \frac{\partial f}{\partial t}$ in $H_D$ is necessary for the gauge invariance of
the Dirac equation under time-dependent gauge transformation.

Denoting the electromagnetic potential produced by the static proton source as
$A^\mu=(A^0, {\vec A})=(\frac{e}{4\pi r}-\frac{\partial f}{\partial t}, {\vec \nabla}f)$,
one can write
\begin{equation}
H = {\vec \alpha} \cdot ({\vec p}+e{\vec A})+\beta m-e A^0_{phys}
\end{equation}
and
\begin{equation}
H_D = {\vec \alpha} \cdot ({\vec p}+e{\vec A})+\beta m -eA^0.
\end{equation}
Thus we have explicitly verified that the gauge
invariant hydrogen atom Hamiltonian is different from the Dirac Hamiltonian.

Here we shall give some further discussion on the Dirac Hamiltonian
and the gauge invariant Hamiltonian. In our approach we use the
Dirac Hamiltonian in the time-dependent Dirac equation, because the
Dirac Hamiltonian is the time translation generator. The gauge
invariant Hamiltonian is the energy operator of the system, whose
eigenvalue is the energy of the hydrogen atom. The gauge invariant
Hamiltonian is generally time-dependent. In this case, one can solve
the energy eigenvalue equation at any specific instant of time. It
can be shown that the energy eigenvalues are gauge independent, and
by suitably choosing the phase factor of the time-dependent
eigenfunction, one can ensure that the time-dependent eigenfunction
satisfies the Dirac equation. The proof is as follows.

Let $H_C$ be the gauge invariant Hamiltonian in the Coulomb gauge
\begin{equation}
H_C={\vec \alpha}\cdot {\vec p}+\beta m -\frac{e^2}{4\pi r},
\end{equation}
and $H_f(t)$ be the gauge invariant Hamiltonian in a general gauge
\begin{equation}
H_f(t)={\vec \alpha}\cdot ({\vec p}+e {\vec \nabla}f)+\beta m
-\frac{e^2}{4\pi r}.
\end{equation}
It can be easily seen that $H_f(t)$ and $H_C$ are connected by a
time-dependent unitary transformation:
\begin{equation}\label{unitarytransform}
e^{-ief({\vec x},t)}H_C e^{ief({\vec x},t)}=H_f (t).
\end{equation}
$H_C$ has the following energy eigenvalue equation
\begin{equation}\label{eigenvalueEq}
H_C \big(e^{-iE_n t}\psi_n({\vec x})\big)=E_n \big(e^{-iE_n
t}\psi_n({\vec x})\big),
\end{equation}
where $E_n$ is the energy eigenvalue of the hydrogen atom in the
Coulomb gauge and $e^{-iE_n t}\psi_n({\vec x})$ is the corresponding
stationary state wave function. From (\ref{unitarytransform}) and
(\ref{eigenvalueEq}) one can derive
\begin{equation}
H_f(t)\big(e^{-ief({\vec x},t)}e^{-iE_n t}\psi_n({\vec
x})\big)=E_n\big( e^{-ief({\vec x},t)}e^{-iE_n t}\psi_n({\vec
x})\big).
\end{equation}
So, at each instant of time $t$, $H_f(t)$ has the same eigenvalues
as the Coulomb gauge Hamiltonian with $e^{-ief({\vec x},t)}e^{-iE_n
t}\psi_n({\vec x})$ being the corresponding instantaneous
eigenfunction. In addition, since $e^{-iE_n t}\psi_n({\vec x})$
satisfies the time-dependent Dirac equation in the Coulomb gauge
\begin{equation}
i\frac{\partial}{\partial t}\big(e^{-iE_n t}\psi_n({\vec
x})\big)=({\vec \alpha}\cdot {\vec p}+\beta m -\frac{e^2}{4\pi
r})\big(e^{-iE_n t}\psi_n({\vec x})\big),
\end{equation}
from the gauge invariance of the Dirac equation, one has
\begin{equation}
i\frac{\partial}{\partial t} \big( e^{-ief}e^{-iE_n t} \psi_n({\vec
x})\big)=\bigg({\vec \alpha}\cdot ({\vec p}+e {\vec \nabla}f)+\beta
m -\frac{e^2}{4\pi r}+e\frac{\partial f}{\partial t} \bigg)\big(
e^{-ief}e^{-iE_n t} \psi_n({\vec x})\big).
\end{equation}
Therefore, the instantaneous eigenfunction $e^{-ief}e^{-iE_n t}
\psi_n({\vec x})$ of the time-dependent gauge invariant Hamiltonian
$H_f(t)$ satisfies the time-dependent Dirac equation.

\section{IV.~SUMMARY}

Gauge invariance has long been recognized as a first principle
through the development of the standard model. However, for quantum
mechanics of a charged particle in a classical external
electromagnetic field, there is an apparent puzzle that the matrix
element of the canonical momentum and Hamiltonian operators  is
gauge dependent. A resolution to this puzzle is provided by us in
\cite{ChenWang}. Based on the separation of the electromagnetic
potential into pure gauge and gauge invariant parts, we have
proposed a new set of momentum and Hamiltonian operators which
satisfy both the requirement of gauge invariance and the relevant
commutation relations.

In this paper we did a check for the case of the hydrogen atom
problem: Starting from the Hamiltonian of the coupled electron,
proton and electromagnetic field, under the infinite proton mass
approximation, we derive the gauge invariant hydrogen atom
Hamiltonian and verify explicitly that this Hamiltonian is different
from the Dirac Hamiltonian, which is the time translation generator
of the system. Therefore, the Dirac Hamiltonian, which determines
the time evolution of the system, is not an observable, whereas the
gauge invariant Hamiltonian is. The gauge invariant Hamiltonian is
the energy operator, whose eigenvalue is the energy of the hydrogen
atom. It is generally time-dependent. In this case, one can solve
the energy eigenvalue equation at any specific instant of time. It
is shown that the energy eigenvalues are gauge independent, and by
suitably choosing the phase factor of the time-dependent
eigenfunction, one can ensure that the time-dependent eigenfunction
satisfies the Dirac equation.

\end{document}